\documentclass[]{aa}
\usepackage{graphicx}
\usepackage{txfonts}
\usepackage{psfig}

\begin{document}

\title{Detection of strongly processed ice in the central starburst of NGC\,4945
       \thanks{Based on observations collected at the European Southern 
       Observatory, Chile in programs 63.N--0387, 65.N--0535\,\&\,67.B--0096}}

\author{H.W.W. Spoon\inst{1}
       \and
        A.F.M. Moorwood\inst{2}
       \and
        K.M. Pontoppidan\inst{3}
       \and
        J. Cami\inst{4}
       \and
        M. Kregel\inst{1}
       \and
        D. Lutz\inst{5}
       \and
        A.G.G.M. Tielens\inst{1,6}
       }

\offprints{H.W.W. Spoon ({\tt spoon@astro.rug.nl})}

\institute{Kapteyn Institute, P.O. Box 800, NL-9700 AV Groningen, 
           The Netherlands
          \and
           European Southern Observatory, Karl-Schwarzschild-Strasse 2,
           D-85748 Garching, Germany
          \and
           Leiden Observatory, P.O. Box 9513, NL-2300 RA Leiden,
           The Netherlands
          \and
           NASA-Ames Research Center, Mail Stop 245-6, Moffett Field, 
           CA 94035, USA
          \and
           Max-Planck-Institut f\"ur Extraterrestrische Physik (MPE),
           P.O. Box 1312, D-85741 Garching, Germany
          \and
           SRON, P.O. Box 800, NL-9700 AV Groningen, The Netherlands
          }

\date{Received date; accepted date}

\abstract{The composition of ice grains provides an important tool for
the study of the molecular environment of star forming regions. Using 
ISAAC at the VLT to obtain spectra around 4.65\,$\mu$m we have detected 
for the first time `XCN' and CO ice in an extragalactic environment: 
the nuclear region of the nearby dusty starburst/AGN galaxy NGC\,4945.
The profile of the solid CO band reveals the importance of thermal
processing of the ice while the prominence of the XCN band attests to
the importance of energetic processing of the ice by FUV radiation
and/or energetic particles. In analogy to the processing of ices
by embedded protostars in our Galaxy, we attribute the processing 
of the ices in the center of NGC\,4945 to ongoing massive star 
formation. Our M-band spectrum also shows strong HI\,Pf$\beta$
and H$_2$ 0--0 S(9) line emission and gas phase CO absorption lines.
The HI, H$_2$, PAH, gas phase CO and the ices seem to be embedded 
in a rotating molecular disk which is undergoing vigorous star 
formation.
Recently, strong OCN$^-$ absorption has been detected in the spectrum 
of the Galactic center star GC:IRS\,19. The most likely environment for 
the OCN$^-$ absorption is the strongly UV-exposed GC molecular ring.
The presence of processed ice in the center of NGC\,4945 and 
our Galactic center leads us to believe that processed ice may
be a common characteristic of dense molecular material in star 
forming galactic nuclei.

\keywords{Galaxies: individual: NGC4945 ---
                    Galaxies: ISM --- 
                    Galaxies: nuclei --- 
                    Galaxies: starburst --- 
                    Infrared: galaxies }}

\maketitle

\section{Introduction}
It has long been recognized that dust is an important component 
of the ISM in star forming regions in our galaxy and nearby galaxies.
Optical studies of high redshift galaxies as well as deep mid-infrared
and submm surveys have recently stressed the importance of dust in
galaxies up to high redshifts. Understanding
the composition, origin and evolution of dust -- particularly in
star forming regions -- is therefore a key question of astrophysics.
Of special importance are interstellar ices, which are only present
in molecular cloud environments. Mid-infrared spectra taken with the 
spectrometers aboard the Infrared Space Observatory (ISO)
have shown ices to exist in a variety of extragalactic sources, from 
nearby starburst nuclei up to distant dust-enshrouded ultra-luminous
infrared galaxies (ULIRGs). 
Weak H$_2$O ice absorptions were first detected in the nuclear spectra 
of the nearby galaxies M\,82 and NGC\,253 (Sturm et al. \cite{Sturm00}). 
Strong ice absorptions of H$_2$O, CO$_2$ and the XCN/CO blend were 
first seen in the nucleus of NGC\,4945 (Spoon et al. \cite{Spoon00}). 
This was followed by the discovery of H$_2$O, `HAC' and CH$_4$ ice in 
the nuclear spectrum of NGC\,4418 (Spoon et al. \cite{Spoon01}).
So far ices have been found in some twenty galaxies (Spoon et al. 
\cite{Spoon02}), the most distant of which is IRAS\,00183--7111,
a ULIRG at z=0.33 (Tran et al. \cite{Tran00}).

Interstellar ices are ideal probes for the conditions in the coldest and 
best shielded galaxy components -- their molecular clouds. Embedded
protostars can process interstellar ice in their environment thermally
as well as through FUV photolysis. Solid CO
and `XCN' provide prime probes for this processing. The fundamental
vibrational modes of these species correspond to wavelengths around 
4.65\,$\mu$m in the M-band atmospheric window.
CO ice is highly sensitive to thermal processing of its environment.
In quiescent molecular clouds, as probed by e.g. the field star Elias\,16,
the CO ice feature is dominated by the feature characteristic for an
apolar ice mixture. When heated to temperatures above 20\,K, this ice 
mixture sublimates and the only CO ice mixture surviving is a
mixture dominated by H$_2$O ice. This mixture has a distinctly
different profile and sublimates at temperatures of $\sim$90\,K. The
spectrum of the deeply embedded massive protostar W\,33A is dominated 
by this type of CO profile (Chiar et al. \cite{Chiar98}). The presence 
of `XCN', or OCN$^{-}$ ice, after its most popular identification 
(Demyk et al. \cite{Demyk98}), is indicative of strong processing of 
icy grains by UV light, energetic particles or heating 
(Lacy et al. \cite{Lacy84}; Grim et al. \cite{Grim89}; 
Palumbo et al. \cite{Palumbo00}; F. van Broekhuizen, priv. comm.). 
In our Galaxy, OCN$^{-}$ ice is strongest in the embedded massive 
protostar W\,33A (e.g. Pendleton et al. \cite{Pendleton99}). 
The feature is generally weak (compared to CO ice) in 
other protostars (e.g. Pendleton et al. \cite{Pendleton99}) and absent in 
quiescent molecular clouds, such as probed by the line of sight towards 
the field star Elias\,16 (Chiar et al. \cite{Chiar95}).

\begin{figure*}[!thb]\centering
\includegraphics[width=10cm,angle=-90]{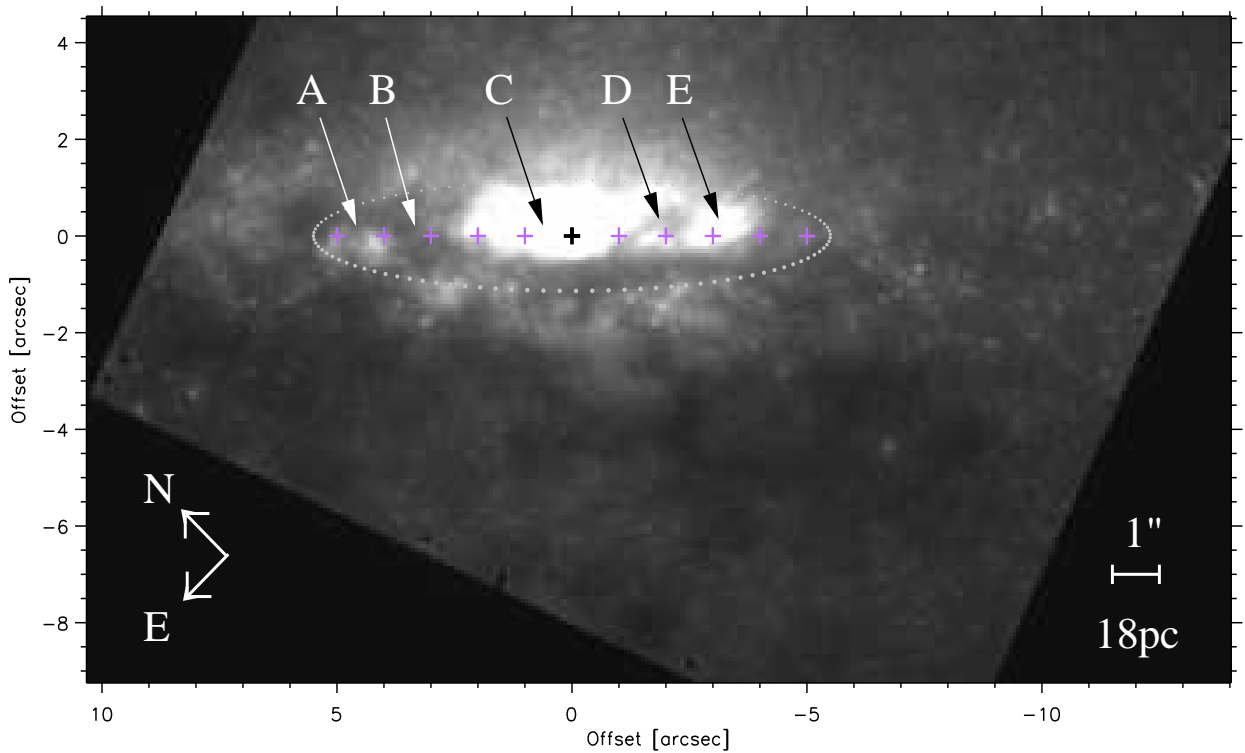}
\caption{HST/NICMOS K-band image of the nucleus of NGC\,4945 (reproduced
from Marconi et al. (\cite{Marconi00}). The image has been rotated so
that the galaxy major axis (PA=43$\degr$) is oriented horizontally.
The position of the VLT/ISAAC slit is marked by crosses, spaced 
1$\arcsec$ apart. Arrows indicate several positions of interest, 
which are discussed in the text. The dotted ring gives an impression
of how a circumnuclear ring with radius 5.5$\arcsec$ (100\,pc) 
would be seen under an inclination of 78$\degr$.}
\label{KBandIma}
\end{figure*}

Here we report on our VLT/ISAAC L and M-band follow-up spectroscopy of 
the rich ice absorption spectrum of the nucleus of NGC\,4945 (Spoon et al.
\cite{Spoon00}), a nearby (3.9\,Mpc; 1$\arcsec$=18\,pc; 
Bergman et al. \cite{Bergman92}) luminous 
(L$_{\rm IR}$=3$\times$10$^{10}$\,L$_{\odot}$) 
infrared galaxy, seen nearly edge-on (i$\sim$78$\degr$; 
Ables et al. \cite{Ables87}).
The central region of this galaxy is dominated by a visually obscured 
starburst (Moorwood et al. \cite{Moorwood96}; Marconi et al. \cite{Marconi00}) 
and a heavily enshrouded AGN, only seen in hard X-rays 
(Iwasawa et al. \cite{Iwasawa93}; Guainazzi et al. \cite{Guainazzi00}). 
Pa$\alpha$ (1.88\,$\mu$m) and K-band (2.2\,$\mu$m) images obtained with 
HST/NICMOS (Marconi et al. \cite{Marconi00}) reveal a complicated nuclear 
morphology, resulting from a nuclear starburst partially obscured by a 
strongly absorbing circumnuclear star forming ring, seen nearly edge-on.
Fig.\,\ref{KBandIma} shows the HST/NICMOS K-band image
(Marconi et al. \cite{Marconi00}), rotated such that the galaxy major axis 
lies horizontally. A possible geometry for the circumnuclear starburst
ring, seen under an inclination of $\sim$78$\degr$, is indicated by a 
dotted circle with radius 5.5$\arcsec$ (100\,pc). The apparent asymmetric
distribution of the nuclear emission interior to this ring (strong emission 
at position E, no emission at position B) has been attributed to patchy 
absorption within the ring (Marconi et al. \cite{Marconi00}). Note that 
the emission at position A on the galaxy major axis appears to arise in 
the circumnuclear ring, not in the nuclear starburst.
The dark structures seen at position D in Fig.\,\ref{KBandIma} are likely 
gas filaments rising above the star forming ring and appearing in 
absorption against the bright nuclear continuum. The Pa$\alpha$ image
(Marconi et al. \cite{Marconi00}) shows more of these dark filamentary 
structures, one of which passes in front of the K-band nucleus.
The AGN, the second brightest Seyfert 2 nucleus in the sky at hard X-rays 
(Done et al. \cite{Done96}), does not appear in any of the HST/NICMOS
images. Groundbased 10\,$\mu$m ESO/TIMMI images (1$\arcsec$ seeing) 
reveal no sign of the central monster either (Marconi et al. \cite{Marconi00}).
Also mid-infrared spectroscopic AGN tracers, like 7.65\,$\mu$m 
$[\ion{Ne}{vi}]$ and 14.3\,\&\,24.3\,$\mu$m $[\ion{Ne}{v}]$, resulted in 
strong upper limits (Spoon et al. \cite{Spoon00}). 
It is hence likely that the AGN is strongly obscured in all directions
by material which is most probably close to the black hole for the 
obscuration to be effective. The conical cavity, protruding from the 
nucleus along the minor axis and seen in the near-infrared 
(Moorwood et al. \cite{Moorwood96}) and in soft X-rays (Schurch et al.
\cite{Schurch02}), is therefore not a Seyfert ionization cone, but a 
cavity cleared by a supernova-driven starburst 'superwind' 
(Moorwood et al. \cite{Moorwood96}).

Here we present infrared L and M-band spectra of all nuclear components
discussed above obtained using VLT/ISAAC with its slit aligned with the
galaxy major axis (Fig.\,\ref{KBandIma}).

\section{Observations}

The 3--5\,$\mu$m spectra of the NGC\,4945 nucleus were obtained using the 
Infrared Spectrometer And Array Camera (ISAAC) mounted on the VLT (UT1-Antu) 
at the ESO Paranal Observatory. The observations were performed on the night 
of June 15, 2000 for the L-band spectrum and on the nights of June 15--16, 2000 
and July 1--3, 2001 for the M-band spectrum. The low resolution grating and 
the 1$\arcsec$ slit were used for the L-band spectrum with a corresponding 
resolving power of R=360. The M-band spectrum was obtained using the medium 
resolution grating and the 1$\arcsec$ slit, resulting in a spectral resolving 
power of  R=3000 ($\Delta$v=100\,km/s FWHM). In both cases, a single 
setting was done covering the entire L-band and the region from 4.56 to 
4.80\,$\mu$m in the M-band. The spectra have a total integration time of 
240\,s and 5123\,s for the L and the M-band, respectively. 
The telescope was operated in the standard chop-nod mode with a 
chop throw of 20$\arcsec$. The telluric standard BS\,5571 (B2III) was observed
just before or after each observation of NGC\,4945 with a maximal airmass 
difference of 0.15.

The data were reduced using our own IDL routines. The individual frames were 
corrected for the non-linearity of the Aladdin detector, distortion corrected 
using a star trace map and bad pixels and cosmic ray hits were removed before 
co-adding. For the M-band this procedure resulted in six co-added source 
frames and six co-added standard star frames. Standard star spectra were 
obtained by extracting the positive spectral trace from each of the six 
standard star frames. The six source frames were then divided by their 
associated standard star spectra, taking into account an optimal small shift 
between the source frame and standard spectrum by requiring that the 
pixel-to-pixel noise on the continuum of the final source frame be 
minimized. No correction for airmass differences was attempted due to 
insufficient signal-to-noise ratio of the source frames. No hydrogen 
recombination lines were detected in the standard spectrum and thus no 
attempts were made to correct the standard for photospheric lines.
Next, each source frame was flux calibrated relative to the standard and 
wavelength calibrated using the telluric absorption lines in the standard 
star spectrum. The final source frame was then produced by stacking the 
six source frames, taking into account small dispersion shifts among the 
frames.

The flux calibration is estimated to be better than 15\% and the wavelength 
calibration is accurate to 150\,km/s and 15\,km/s for the L and M-band 
spectra, respectively.

\section{Results}

\subsection{The 3\,$\mu$m water ice band}

\begin{figure}[]
 \begin{center}
  \psfig{figure=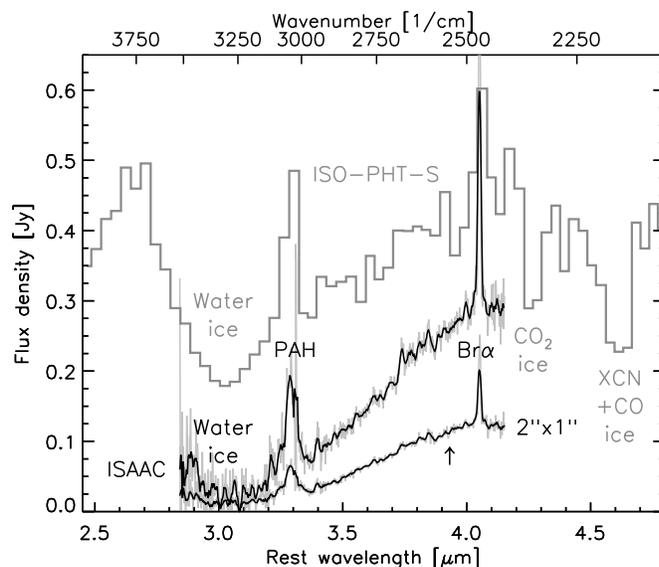,angle=0}
 \end{center}
\vspace*{-10mm}
\caption{The VLT/ISAAC rest frame L-band spectrum of the central 
2$\arcsec\times$1$\arcsec$ (36\,$\times$\,18\,pc$^2$) and
11.5$\arcsec\times$1$\arcsec$ (210\,$\times$\,18\,pc$^2$) of 
NGC\,4945 (black) compared with the ISO--PHT--S spectrum obtained in a 
24$\arcsec\times$24$\arcsec$ aperture (grey histogram). The arrow
indicates the expected wavelength of the 3.93\,$\mu$m $[\ion{Si}{ix}]$
line.}
\label{LBandFig}
\end{figure}

The ISAAC L-band spectrum of the central region of NGC\,4945 
(Fig.\,\ref{LBandFig}) is dominated by a broad (2.7--4.0\,$\mu$m) 
absorption band attributed mainly to water ice 
(e.g. Smith et al. \cite{Smith89}; Chiar et al. \cite{Chiar00}). The
width of the feature is best appreciated in the low resolution 
24$\arcsec\times$24$\arcsec$ ISO-PHT-S spectrum, which is shown for 
comparison. 
Superimposed on the absorbed L-band continuum are the PAH emission 
bands at 3.3\,\&\,3.4\,$\mu$m as well as the H\,Br$\alpha$ line at 4.05\,$\mu$m. 
The ISAAC spectra of the central 2$\arcsec\times$1$\arcsec$ and 
11.5$\arcsec\times$1$\arcsec$ agree well with the larger beam 
ISO-PHT-S spectrum, assuming the water ice feature in the latter 
spectrum to be diluted by a non-absorbed stellar and dust continuum.
Adopting a flat continuum (fixed at 4.0--4.1\,$\mu$m) and depending on 
whether we fit the bottom or the blue wing of the feature, we find 
a water ice column of 41--47\,$\times$\,10$^{17}$\,cm$^{-2}$ (assuming
a band strength of 2.0\,$\times$10$^{16}$\,cm/molecule; 
Gerakines et al. \cite{Gerakines95}).
Note the absence of 3.94\,$\mu$m $[\ion{Si}{ix}]$ line emission from
the ISAAC L-band spectrum. This line arises in soft X-ray photoionized
gas and is comparable to or brighter than the H Br$\alpha$
line in many Seyfert galaxies (Oliva et al. \cite{Oliva94}; Lutz
et al. \cite{Lutz02}). The absence ($[\ion{Si}{ix}]$/H\,Br$\alpha<$0.10) 
may be taken as evidence for the extremely high obscuration of the 
AGN in NGC\,4945 in the L-band. 

\subsection{Processed CO and OCN$^-$ ice}

\begin{figure}[]
 \begin{center}
  \psfig{figure=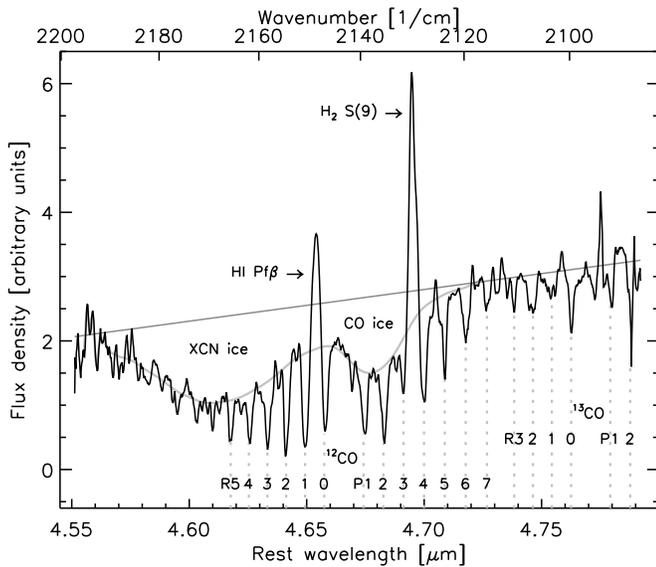,angle=0}
 \end{center}
\vspace*{-10mm}
\caption{The VLT/ISAAC rest frame M-band spectrum of the central 
2$\arcsec\times$1$\arcsec$ (36\,$\times$\,18\,pc$^2$) of NGC\,4945.
The spectrum shows emission lines of HI\,Pf$\beta$ and H$_2$ 0--0 S(9),
broad absorption features of ices containing XCN and CO, and
absorption lines of gas phase $^{12}$CO and $^{13}$CO. The adopted 
continuum and the shape of the ice feature, corrected for the presence 
of gas phase $^{12}$CO and $^{13}$CO lines, are drawn in grey.}
\label{MBandFig}
\end{figure}

The M-band spectrum of the central 2$\arcsec\times$1$\arcsec$ of the 
nuclear region of NGC\,4945 (Fig.\,\ref{MBandFig}) contains strong and
relatively broad absorption features at 4.62 and 4.67\,$\mu$m as well 
as gas lines due to H, H$_2$ and CO. The line centers of the gas phase 
lines shift as a function of position along the slit.

\begin{figure}[]
 \begin{center}
  \psfig{figure=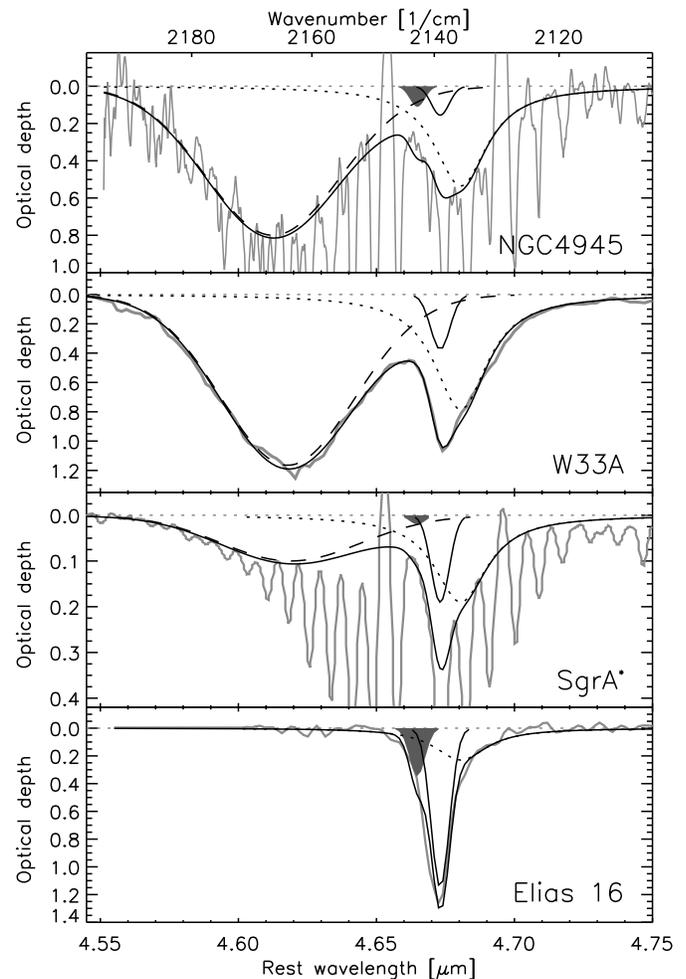,angle=0}
 \end{center}
\vspace*{-10mm}
\caption{The optical depth spectrum of ices in NGC\,4945 compared to
ices seen in Galactic lines of sight. The {\bf Top panel} shows the 
optical depth spectrum of the central 2$\arcsec\times$1$\arcsec$ 
(36\,$\times$\,18\,pc$^2$) of NGC\,4945 (grey), determined adopting 
the continuum shown in Fig.\,\ref{MBandFig}. Also indicated are 
the results of the fit to the ice profile. The OCN$^-$ fit component 
is denoted by a dashed line, the polar CO component by a dotted line, 
the pure CO component by a continuous line and the apolar component 
by a grey surface. The overall fit is shown as a black line. The 
spectrum was obtained at a resolving power of R=3000.
{\bf Second panel:} Optical depth spectrum for the massive embedded 
protostar W\,33A. The spectrum has been degraded to a resolving power 
of R=750.
{\bf Third panel:} Idem for the ISO-SWS line of sight to the Galactic 
Center (Sgr\,A$^*$). The spectrum was obtained at a resolving power 
of R=2000.
{\bf Bottom panel:} Idem for the unprocessed line of sight to field 
star Elias\,16, located behind the Taurus molecular cloud. The 
spectrum was obtained at a resolving power of R=1200.}
\label{OptDepthFig}
\end{figure}

In order to investigate the various M-band absorption features, we fitted 
a linear continuum to the two pivot ranges 4.54--4.55\,$\mu$m and 
4.78--4.79\,$\mu$m. The resulting optical depth spectrum is shown in 
Fig.\,\ref{OptDepthFig}a. In Fig.\,\ref{OptDepthFig}b, \ref{OptDepthFig}c
and \ref{OptDepthFig}d we show spectra of three comparison objects: the 
embedded massive protostar W\,33A with the strongest known XCN feature; 
the Galactic Center (Moneti et al. \cite{Moneti01}) showing XCN absorption 
towards its massive star forming region which is obscured in what is 
effectively an `edge-on' view towards the center of our own galaxy; 
and, finally, the unprocessed line of sight to the field star Elias\,16 
located behind the Taurus molecular cloud (Chiar et al. \cite{Chiar95}). 
The similarity of NGC\,4945 to W\,33A is particularly striking and will
be discussed later. 

Following Galactic ice studies (Tielens et al.
\cite{Tielens91}; Schutte \& Greenberg \cite{Schutte97}; Chiar et al. 
\cite{Chiar98}; Demyk et al. \cite{Demyk98}; Pendleton et al. 
\cite{Pendleton99}) we identify the prominent absorption feature centered 
at 2168.0\,cm$^{-1}$ (FWHM=24.0\,cm$^{-1}$) with solid state `XCN', or 
OCN$^{-}$ after its most likely identification (Demyk et al. \cite{Demyk98}). 
In order to measure the shape and depth of the feature, we adopt the ice 
feature shown in Fig.\,\ref{MBandFig}.
This continuum has been corrected for the presence of gas phase $^{12}$CO 
and $^{13}$CO absorption lines, discussed later in this Section. 
The OCN$^{-}$ feature appears slightly blueshifted and narrower than in 
W\,33A ($\nu_0$=2165.5\,cm$^{-1}$; FWHM=26.7\,cm$^{-1}$). 
The differences are similar to those found between the embedded 
protostars AFGL\,961 and W\,33A (Pendleton et al. \cite{Pendleton99})
and may be attributed to differences in ice composition 
(Grim \& Greenberg \cite{Grim87}). 
The column density of OCN$^{-}$ is listed in Table\,\ref{OptDepthTab} 
and were computed assuming an OCN$^{-}$ bandstrength of 
1.3$\times$10$^{-16}$ (F. van Broekhuizen, priv. comm.).

Detailed studies of the observed profiles of the solid CO ice band in
Galactic sources have shown that they can all be fitted by varying
contributions of the same three components 
(Boogert et al. \cite{Boogert02a},\,\cite{Boogert02b}; Pontoppidan et al.  
\cite{Pontoppidan03}).
These three components are thought to represent ``pure'' solid CO 
(Gaussian centered at $\nu_0$=2139.9\,cm$^{-1}$ with FWHM=3.5\,cm$^{-1}$),
traces of CO trapped in H$_2$O ice (``polar CO''; Lorentzian centered at
$\nu_0$=2136.5\,cm$^{-1}$ with FWHM=10.6\,cm$^{-1}$) and CO in solid 
CO$_2$ ice (``apolar CO''; Gaussian centered at 
$\nu_0$=2143.7\,cm$^{-1}$ with FWHM=3.0\,cm$^{-1}$)  
(Boogert et al. \cite{Boogert02a}; Pontoppidan et al. \cite{Pontoppidan03}).
Here we note that the spectrum of NGC\,4945 is dominated by traces of CO 
trapped in H$_2$O ice with little or no evidence for the other two 
components (Fig.\,\ref{OptDepthFig}a). In contrast, the spectrum of 
W\,33A (Fig.\,\ref{OptDepthFig}b) has a noticable contribution of the 
``pure'' solid CO component (Tielens et al. \cite{Tielens91}). 
Table\,\ref{OptDepthTab} lists the column densities of the various CO 
ice components, which were computed assuming a CO bandstrength of 
1.1\,$\times$10$^{17}$ cm/molecule (Gerakines et al. \cite{Gerakines95}).

\begin{figure}[]
 \begin{center}
  \psfig{figure=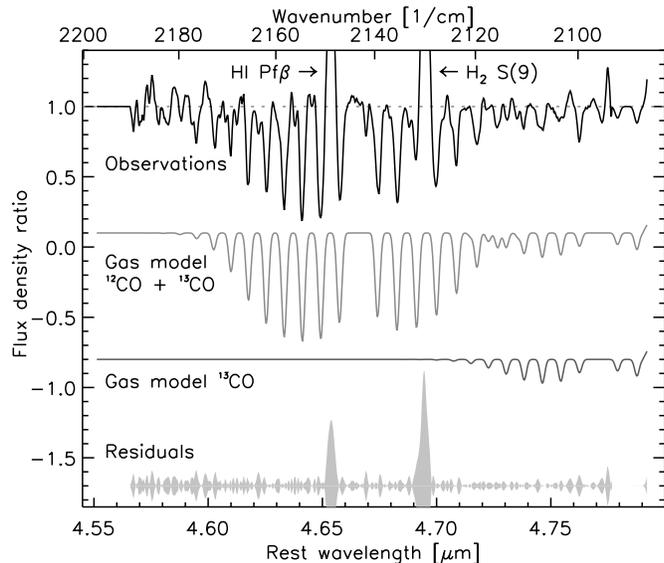,angle=0}
 \end{center}
\vspace*{-10mm}
\caption{CO gas absorption lines in the spectrum of the central 
2$\arcsec\times$1$\arcsec$ (36\,$\times$\,18\,pc$^2$) of NGC\,4945.
The observed spectrum, after dividing out the ice feature, is shown
in black. The best fitting CO gas model is drawn in dark grey. Fit 
residuals are shown in light grey.}
\label{CoGasFig}
\end{figure}

The nuclear spectrum of NGC\,4945 also reveals CO in the gas phase. 
In Fig.\,\ref{MBandFig} we identify a total of 13 fundamental 
ro-vibrational absorption lines of $^{12}$CO and several of $^{13}$CO. 
This is a sufficient number of lines to attempt a single component
model fit to determine the temperature, intrinsic line width and 
$^{12}$CO and $^{13}$CO gas column densities. For this purpose 
we use the isothermal plane-parallel LTE CO gas models of 
Cami et al. (\cite{Cami02}), folded with the appropriate VLT/ISAAC 
spectral resolution (R=3000; $\Delta$v=100\,km/s FWHM). 
In the fitting procedure both the observed 
spectrum and the model spectra are normalized, through division 
by a strongly smoothed version of the respective spectra. 
A four-parameter least-squares minimalization procedure then picks
the best fitting model. Fig.\,\ref{CoGasFig} shows the result for
the nuclear spectrum of NGC\,4945. The CO gas appears moderately
warm at 35$^{+7.5}_{-2.5}$\,K, with an intrinsic line width (FWHM)
of 50$\pm$5\,km/s and column densities 
log N($^{12}$CO)=18.3$\pm$0.1\,cm$^{-2}$ and 
log N($^{13}$CO)=17.2$^{+0.15}_{-0.05}$\,cm$^{-2}$, assuming a
covering factor 1 for the absorber. The uncertainties listed above
do not take into account the systematic errors resulting from the 
data reduction, which may be appreciable. The very low 
$^{12}$CO/$^{13}$CO ratio of $\sim$13 indicates that some $^{12}$CO 
lines are likely to be optically thick. In addition, the absorbing 
material may not be in a uniform screen 'covering factor 1' 
configuration (Fig.\,\ref{KBandIma}). This will particularly affect 
the derived column for $^{12}$CO. A more realistic, yet more 
uncertain, value for the $^{12}$CO gas column may be derived from 
the $^{13}$CO column. Assuming a $^{12}$CO/$^{13}$CO ratio of 80,
we find log N($^{12}$CO)=19.1$^{+0.15}_{-0.05}$\,cm$^{-2}$. 
Note that the (high resolution) spectrum of the physically closest 
resembling object, the protostar W\,33A, also contains CO gas phase 
lines (Mitchell et al. \cite{Mitchell88}). At the resolution of the 
ISO--SWS spectrum (R$\sim$750; Gibb et al. \cite{Gibb00}) these 
are however not detectable.

\subsection{The spatial distribution of the ice}

\begin{figure*}[]
 \begin{center}
  \psfig{figure=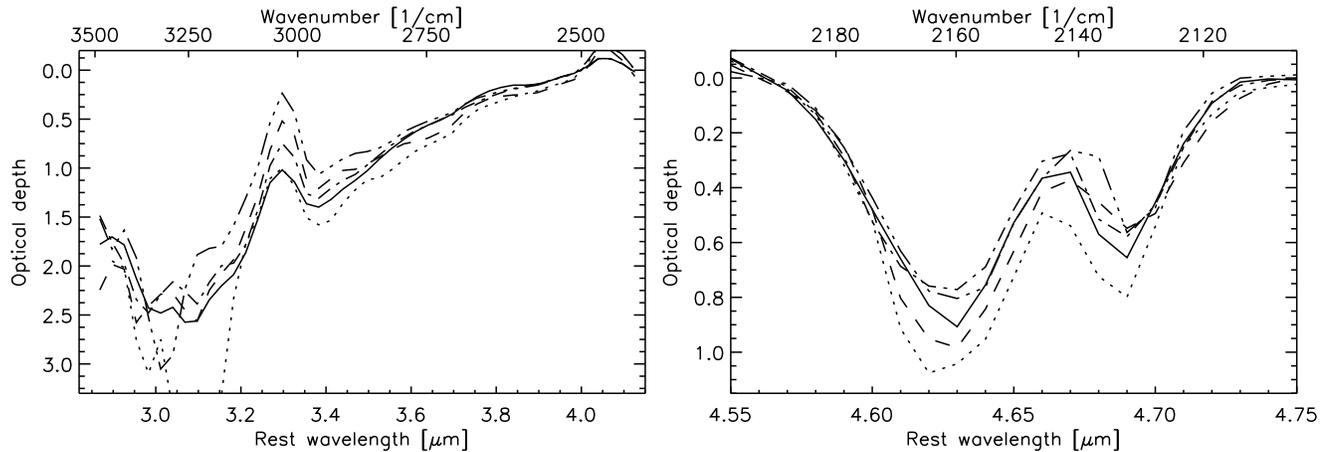,angle=0}
 \end{center}
\vspace*{-10mm}
\caption{Optical depth spectra of ice features in NGC\,4945 for 5 
adjacent positions (3.2$\arcsec$\,SW: {\it triple dot-dashed}, 
1.7$\arcsec$\,SW: {\it dotted}, 0.45$\arcsec$\,SW: {\it solid}, 
0.60$\arcsec$\,NE: {\it dot-dashed}, 1.6$\arcsec$\,NE: {\it dashed}) 
along the galaxy major axis. The 
spectra have been smoothed and rebinned to lower spectral resolution. 
The optical depth spectra reveal no significant differences, except for
the generally larger optical depth at position D, 1.7$\arcsec$\,SW
of the nucleus (dotted line). 
{\bf Left:}  Optical depth spectra of water ice. The feature centered
at 3.3\,$\mu$m is due to the presence of PAH emission in the water ice
absorption feature. {\bf Right:} Optical depth spectra of XCN and CO ice.}
\label{IceBinsFig}
\end{figure*}

\begin{figure}[]
 \begin{center}
  \psfig{figure=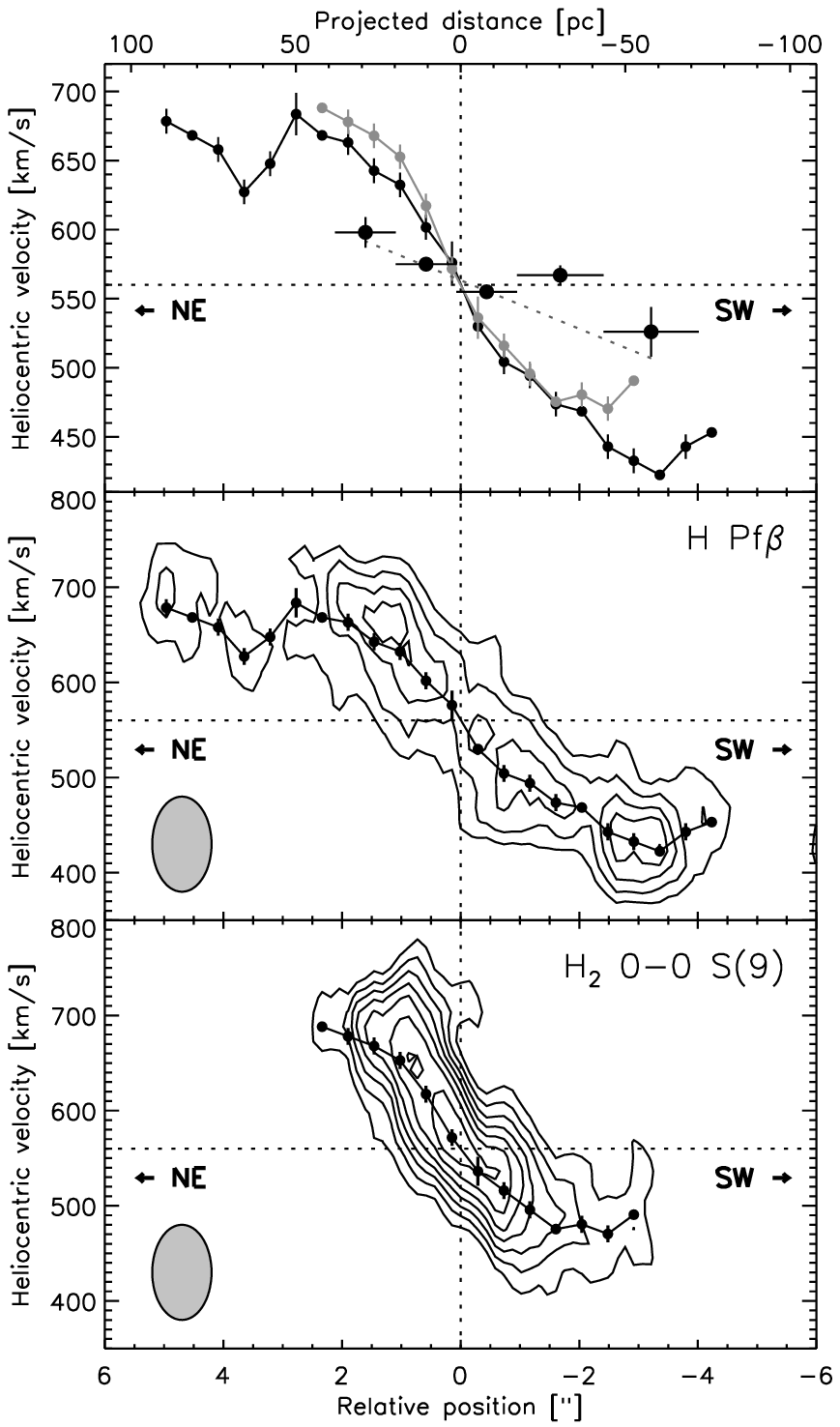,angle=0}
 \end{center}
\vspace*{-10mm}
\caption{Position-velocity information for several species detected in the 
M-band spectrum of NGC\,4945. {\bf Top panel:} Velocity curves for 
HI\,Pf$\beta$ (black) and H$_2$ 0--0 S(9) (grey) as determined from the
position-velocity diagrams in the panels below. The data have been 
rebinned to half the size of a seeing element (0.5$\arcsec$). The five 
large dots denote the CO gas line velocities in five positional 
intervals along the galaxy major axis. A linear fit to the CO gas line
points (excluding the point at 1.7$\arcsec$ SW, which seems to trace
a separate filament) is shown as a dashed line. 
The position of the nucleus is indicated by the vertical dotted line, 
the systemic velocity (561\,km/s) by the horizontal line. 
{\bf Middle panel:} Position-velocity diagram along the major axis for the
HI\,Pf$\beta$ emission line. Contours run from 4 to 10\,$\sigma$ in steps 
of 2\,$\sigma$. The black points indicate the center of the gauss fit to 
the line profile at indicated positions along the galaxy major axis. The 
spatial and spectral resolution is indicated in the lower left corner.
{\bf Lower panel:} Idem for the H$_2$ 0--0 S(9) emission line. Contours run 
from 4 to 14\,$\sigma$ in steps of 2\,$\sigma$.}
\label{RotationCurvesFig}
\end{figure}

The analysis presented above focusses on the nuclear spectrum of NGC\,4945,
which comprises the central 2$\arcsec\times$1$\arcsec$. The nuclear 
continuum emission extends however from $\sim$4.5$\arcsec$ SW to 
$\sim$2$\arcsec$ NE of the nucleus, measured in a 1$\arcsec$ wide slit 
along the SW-NE oriented (PA=43$\degr$) galaxy major axis
(see Fig.\,\ref{KBandIma}). Line emission (Fig.\,\ref{RotationCurvesFig})
can be traced as far out as $\sim$4.5$\arcsec$ SW (position E) and 
$\sim$5.5$\arcsec$ NE (position A) of the nucleus. Both the continuum and line 
emission appear clearly weakened in a $\sim$1$\arcsec$ long strip located 
1.7$\arcsec$ SW (position D) of the nucleus. This strip coincides 
with a dark filamentary structure in the HST/NICMOS K-band image 
(Fig.\,\ref{KBandIma}) of Marconi et al. (\cite{Marconi00}). The same 
image further shows strong extinction from what might be a circumnuclear 
ring to be responsible for limiting the observable part of the major axis 
continuum to the range found (see Sect.\,1). Our analysis further shows 
that the ice and gas absorption features are detected wherever there is
background continuum to absorb. The depth of the ice features appears
to be quite constant (Fig.\,\ref{IceBinsFig}), except for position D, 
where the solid state absorptions may be significantly stronger. Again 
interestingly, this position coincides with the dark filamentary structure 
seen in the HST/NICMOS K-band image (Fig.\,\ref{KBandIma}).

The lower panels of Fig.\,\ref{RotationCurvesFig} show position-velocity 
diagrams for the HI\,Pf$\beta$ and H$_2$ 0--0 S(9) emission lines. The 
distributions are remarkably similar over the central $\pm$2$\arcsec$, 
both indicating rotation about the nucleus. The only significant deviation 
occurs at a position 0.7\,$\arcsec$ NE of the nucleus (position C), 
where the H$_2$ 0--0 S(9) emission extends 
$\sim$40\,km/s beyond the highest velocity traced by HI\,Pf$\beta$. We
speculate that this position may coincide with the start of the N--S 
oriented molecular ridge, traced in H$_2$ 1--0 S(1) by Moorwood et al. 
(\cite{Moorwood96}) and Marconi et al. (\cite{Marconi00}), bordering
the conical cavity cleared by a starburst superwind (Moorwood et al. 
\cite{Moorwood96}). Another deviation occurs 3$\arcsec$--\,4$\arcsec$
NE of the nucleus (position B), where a dark cloud in the circumnuclear 
ring weakens the line emission from the nuclear starburst it eclipses. 
Given the high inclination of the circumnuclear ring, the dark cloud 
may actually be in an orbit in the outer part of the circumnuclear
ring, with an appreciable tangential orbital velocity component. This
would explain why the HI\,Pf$\beta$ emission from that cloud is 
$\sim$50\,km/s lower than that of the bright patch 1$\arcsec$ NE of
it (position A). 
The top panel of Fig.\,\ref{RotationCurvesFig} shows the velocity curves 
derived from fitting single Gaussians to the line emission in each of 
the observed HI\,Pf$\beta$ and H$_2$ 0--0 S(9) position-velocity 
distributions. The velocity curves 
obtained for both emission lines are consistent and yield the same 
velocity gradient of $\sim$60\,km/s/arcsec across the nuclear region. 
The heliocentric systemic velocity of 561$\pm$4\,km/s 
(Dahlem et al. \cite{Dahlem93}) occurs at the brightest M-band cross 
dispersion pixel. Away from the nucleus the velocity curves flatten 
off to rotational velocities 120$\pm$10\,km/s above and below systemic.
This value agrees well with the results for H\,Br$\gamma$ and 
H$_2$ 1--0 S(1) obtained by Moorwood \& Oliva (\cite{Moorwood94}). 
Note that Ott et al. (\cite{Ott01}) find rotational velocities up to
160 km/s for HI (21cm). Assuming that the measured velocities trace the
potential, the mass within the nuclear region amounts to 
1\,$\times$10$^8$\,M$_{\sun}$ at 50\,pc and 3\,$\times$10$^8$\,M$_{\sun}$ 
at 100\,pc. Also shown in the upper panel of Fig.\,\ref{RotationCurvesFig} 
is the velocity curve for gas phase CO. The CO absorption line velocities 
were derived by cross-correlating the CO absorption line spectra with the 
best fitting CO gas model. In order to acquire sufficient signal-to-noise 
in the lines, the spatial information has been combined into 5 bins 
of 1.0$\arcsec$--1.5$\arcsec$ along the major axis. The resulting
velocity curve for the CO absorption lines (excluding the point at
1.7$\arcsec$ SW; position D; see below) clearly samples a different 
velocity field 
than the emission lines do. Based on the four CO points, we measure a 
velocity gradient for the CO gas of 17\,km/s/arcsec across the nucleus.
The velocity measured for the 1.7$\arcsec$ SW point (position D)
clearly does not fit in with the velocity gradient seen for the other 
four CO points. Interestingly, the HST/NICMOS K-band image 
(Fig.\,\ref{KBandIma}) shows at this position a dark filamentary 
structure against the bright K-band continuum. Possibly, the CO gas 
radial velocity measured at 1.7$\arcsec$ SW is related to this 
foreground (circumnuclear ring?) structure instead of to the rotating 
inner ring/disk.

\begin{table}
\caption[]{Measured column densities (10$^{17}$\,molecules/cm$^2$) and 
column density ratios for NGC\,4945, the massive embedded protostar W\,33A, 
the line of sight towards the Galactic center (Sgr\,A$^*$) and towards a 
field star, Elias\,16, located behind the Taurus molecular cloud.}
\label{OptDepthTab}
\begin{tabular}{lllll}
\hline
                        & N4945     & W33A     & Sgr\,A$^*$     & Elias16    \\
\hline
N(polar CO ice)         & 9.4       & 11.4     & 2.8$^e$        & 2.9        \\
N(pure CO ice)          & 0.19      & 1.4      &                & 4.3        \\
N(apolar CO ice)        & 0.16      & 0.0      &                & 1.0        \\
N(total CO ice)         & 9.7       & 12.7     & 3.5$^e$        & 8.2        \\
N(XCN ice)              & 1.6       & 2.6      & 0.2$^e$        & $<$0.13$^b$\\
N(cold CO gas)\,$^a$    & --        & 130$^c$  & 66$^d$         & 18$^h$     \\
N(warm CO gas)\,$^a$    & 130       & 140$^c$  & 1$^d$          & --         \\
N(CO gas)               & 130       & 270$^c$  & 67$^d$         & 18$^h$     \\
N(H$_2$O ice)           & 41--47    & 110$^b$  & 12$^f$         & 25$^g$     \\
N(CO$_2$ ice)         &$\geq$2.0$^i$& 14$^j$   & 1.7$^j$        & 5$^j$      \\
\hline
N(polar CO)/N(CO)      &  0.97      & 0.89     & 0.8            & 0.35       \\
N(XCN)/N(CO)           &  0.17      & 0.20     & 0.07           & $<$0.02    \\
N(CO)/N(H$_2$O)        & 0.21--0.24 & 0.12     & 0.3            & 0.33       \\
N(XCN)/N(H$_2$O)       &0.034--0.039& 0.024    & 0.02           & $<$0.005   \\
N(CO ice)/N(CO gas)    &  0.08      & 0.047    & 0.05           & 0.5        \\
\hline
\end{tabular}


a) We define CO gas as cold if T$_{\rm gas}<$30\,K  and as warm if T$_{\rm gas}\geq$30\,K;
b) Gibb et al. (\cite{Gibb00});
c) Mitchell et al. (\cite{Mitchell88});
d) Moneti et al. (\cite{Moneti01});
e) Based on the Sgr\,A$^*$ spectrum of Moneti et al. (\cite{Moneti01});
f) Chiar et al. (\cite{Chiar00});
g) Chiar et al. (\cite{Chiar95});
h) Whittet et al. (\cite{Whittet89});
i) Spoon et al. (\cite{Spoon00}) obtained in a 24$\arcsec\times$24$\arcsec$ aperture;
j) Gerakines et al. (\cite{Gerakines99})
\end{table}

\section{Discussion}

The wealth of ISM features detected in the VLT/ISAAC L\,\&\,M-band spectra 
of NGC\,4945 allows us for the first time to study simultaneously the 
conditions of ionized hydrogen gas, molecular hydrogen, PAHs, icy grains 
and cold CO gas in the central region of another galaxy.
Combined with the spatial information obtained in a 1$\arcsec$ wide slit 
oriented along the galaxy major axis, our observations give insight in 
the kinematics and spatial location of the different ISM components 
probed.

\begin{figure}[]
 \begin{center}
  \psfig{figure=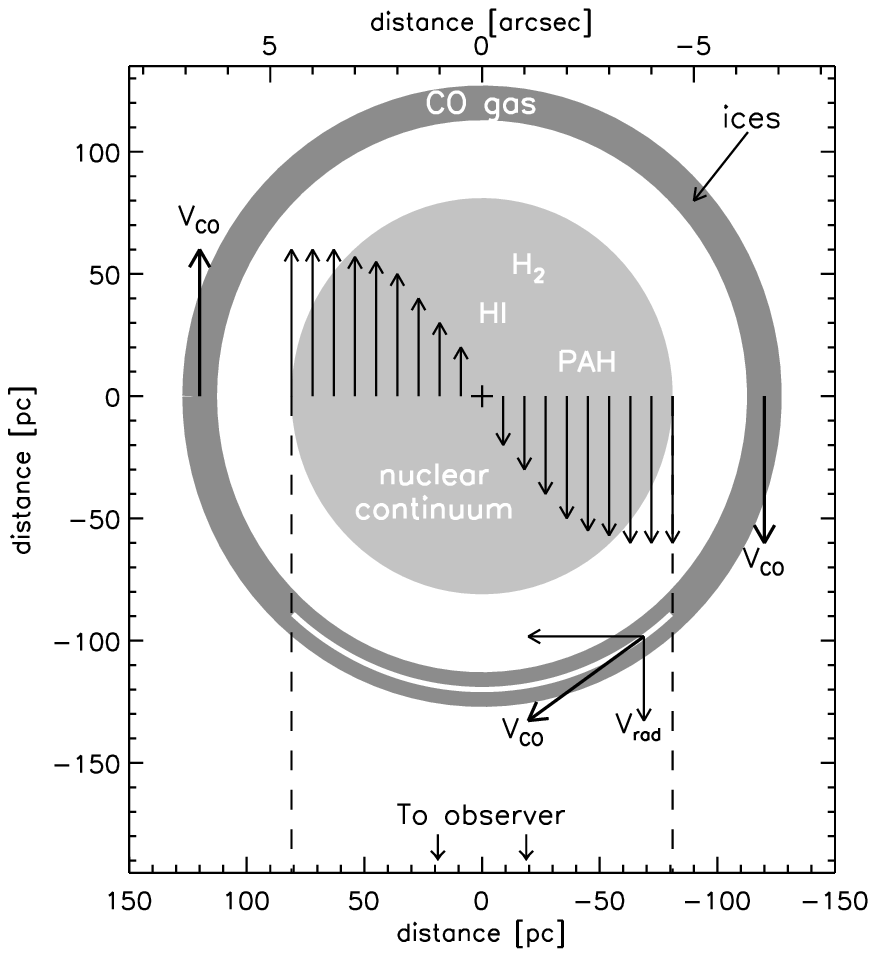,angle=0}
 \end{center}
\vspace*{-10mm}
\caption{Schematic face-on view of the central region of NGC\,4945.
In our model the HI\,Pf$\beta$ and H$_2$\,0--0\,S(9) emission lines
trace the rotation of the inner disk. Their velocity curves 
(Fig.\,\ref{RotationCurvesFig}a) indicate that the rotational velocity 
increases outward to become constant ($\sim$120\,km/s) at r$\sim$60\,pc.
Assuming the CO gas phase absorption lines to arise in the outskirts of
the rotating disk (V$_{\rm rot}$=120\,km/s), the CO velocity gradient 
(17\,km/s/arcsec$\,\simeq$1\,km/s/pc; Fig.\,\ref{RotationCurvesFig}a) 
sampled in front of the nuclear continuum indicates that the ``CO ring'' 
must have a radius of $\sim$120\,pc. The white strip in the CO ring 
indicates the section of the CO ring 
where CO gas absorption lines can be detected against the 
nuclear continuum. The line of sight sight velocity component 
(V$_{\rm rad}$) in this strip changes only slowly with projected 
distance, in agreement with the observed shallow CO velocity curve
(Fig.\,\ref{RotationCurvesFig}a).}
\label{NucGeomFig}
\end{figure}

The picture that emerges is of an extended, fragmented and vigorously 
star forming rotating molecular ring/disk, surrounding the deeply buried, 
and hence passified, AGN. The H, H$_2$ and 
PAHs trace the interaction of the massive stars -- interspersed within 
the ring, as indicated by the velocity behavior of the H and H$_2$ lines 
-- creating HII regions and PDRs. The extended continuum emission source 
created this way, forms the background against which we see the various 
absorption features. The contribution of the AGN to this continuum is
insignificant, as may be concluded from the absence of a pointlike source 
in both the HST K-band image (Marconi et al. \cite{Marconi00}) and our 
ISAAC spectra. Hard X-ray observations have shown that the direct 
view of the AGN is blocked by a hydrogen column density of 10$^{24.7}$ 
cm$^{-2}$ (Iwasawa et al. \cite{Iwasawa93}; A$_{\rm V}\sim$2500). The 
absence of the 3.94\,$\mu$m $[\ion{Si}{ix}]$ coronal line and other 
NLR fine structure lines (Fig.\,\ref{LBandFig}; 
Spoon et al. \cite{Spoon00}) from this region indicates that the 
ionizing radiation does not escape in other directions either.
The shallow velocity curve of the cold CO gas is consistent with the
CO being distributed in the outskirts of the nearly edge-on, rotating 
molecular ring/disk, of which the CO gas velocities are sampled only in 
front of the bright nuclear continuum, where the line of sight velocity 
component increases only slowly with projected distance (see 
Fig.\,\ref{NucGeomFig}). An indication of 
its scale can be obtained in the simplified assumption of absorption in 
a single ring rotating at 120\,km/s. To reproduce the observed velocity 
gradient of CO absorption, the radius of this ring would be 
$\sim$7$\arcsec$ ($\sim$120\,pc). Other distributions of 
absorbing material on similar scales, and in particular similar to the 
dust ring seen in the data of Marconi et al. (\cite{Marconi00}), will 
reproduce the data as well. The material seen in HI absorption by Ott et al.
(\cite{Ott01}) and the rotating central ($<$20$\arcsec$) mm-wave CO 
emission likely belong to the same structure. Although there is no velocity 
information, we are tempted to co-locate the H$_2$O, CO and OCN$^-$ ice 
with the CO gas. First, the solid H$_2$O, CO and OCN$^-$ can be traced as 
far out as the gas phase CO (Fig.\,\ref{IceBinsFig}). Second, the optical 
depth of the solid state features is constant over the spatial range probed 
(except for the 1.7$\arcsec$ SW point in Fig.\,\ref{IceBinsFig}), implying 
co-location of the ices in a foreground position. Third, the filamentary
structure seen at 1.7$\arcsec$ SW (position D) shows both deeper solid
state features (Fig.\,\ref{IceBinsFig}) and a deviant CO gas velocity,
suggesting co-location of the ices with the CO gas. Fourth, the 
absence of the pure CO ice component indicates grains which have been 
thermally processed to above 20\,K. Hence, unlike for dark cloud lines 
of sight in our galaxy (such as Elias\,16 in the Taurus molecular cloud), 
these ice grains are not located in some random, foreground, dark cloud 
along the line of sight, but instead close to a source of thermal
heating. Fifth, the deep OCN$^-$ ice band shows that the ice is heavily 
processed. 
That again locates the ices within regions of massive star formation 
or close to the AGN rather than some foreground material. From the 
spatial extent and uniformity of the OCN$^-$ ice absorption 
(Fig.\,\ref{IceBinsFig}), a location close to the AGN can, however, 
be excluded. The only likely location therefore remain within or close 
to regions of massive star formation in the circumnuclear starburst. 
Indeed, within our own galaxy, the 
massive protostar W\,33A is an extreme example of processed ices 
(Chiar et al. \cite{Chiar98}; Gibb et al. \cite{Gibb00}). Dark cloud 
material in general does not show any evidence for the OCN$^-$
absorption band (Fig.\,\ref{OptDepthFig}d; Whittet et al. \cite{Whittet01}). 
In fact, no other Galactic, luminous protostar shows such a strong 
OCN$^-$ band.

The presence of OCN$^-$ in ice grain mantles is often taken as a sign
of energetic processing by particles or UV photons. Recent experiments 
suggest however that also thermal processing may result in the formation 
of OCN$^-$ (F. van Broekhuizen, priv. comm.). A fouth possibity presents 
itself in the vicinity of an AGN: processing by AGN X-ray photons. 
The AGN in NGC\,4945 is, however, strongly obscured and only hard X-ray
photons manage to escape. Assuming similar obscuration towards the 
r$\sim$100\,pc ice region in NGC\,4945 as towards our line of sight, 
we estimate a hard X-ray (20--100\,keV) flux of $\sim$0.5\,erg/cm$^2$/s 
from the observations of Guainazzi et al. (\cite{Guainazzi00}). 
This is two orders of magnitude 
less than the UV flux in a photon-dominated region next to massive stars 
(e.g. Tielens \& Hollenbach \cite{Tielens85}). The available very hard 
(unabsorbed) AGN X-rays are hence energetically insignificant compared to 
the starburst UV photons. In addition, such hard X-rays may not couple 
efficiently into individual ice grains. Further laboratory studies are 
required to settle these issues.

The starburst in NGC\,4945 is similar in luminosity to the prototypical 
starbursts M\,82 and NGC\,253 which are also located in the nuclear regions 
of almost edge-on galaxies. Comparing ISO spectra of NGC\,4945 
(Spoon et al. \cite{Spoon00}) to ISO data for M\,82 and NGC\,253 
(Sturm et al. \cite{Sturm00}; F\"orster Schreiber et al. \cite{Foerster03}), 
the obscuration of NGC\,4945 is clearly higher towards the ionized medium
(as derived from the 18.71/33.48\,$\mu$m $[\ion{S}{iii}]$ ratio in
the low density limit), towards the PAH emitting medium (8.6\,$\mu$m
and 11.3\,$\mu$m PAH emission features strongly suppressed by 9.7\,$\mu$m
silicate absorption), and in 3\,$\mu$m water ice absorption 
($\tau_{\rm ice}\sim$2.5). 
It remains unclear, however, whether this is just an on average higher 
absolute absorbing column towards the nuclear region of NGC\,4945, as 
perhaps not implausible given also the slightly smaller physical size of 
its starburst, or whether there are differences in the properties of the 
absorbing medium. 
Higher quality M band spectra of M\,82 and NGC\,253 are needed to test 
whether those starbursts also host processed ices as NGC\,4945, but with
columns corresponding to their lower obscuration.

The spectrum of the nearest edge-on galactic nucleus, our Galactic 
center, also shows absorption features due to CO and OCN$^-$ ice.
The spectrum of GC:IRS\,19 (Chiar et al. \cite{Chiar02}) shows the 
features at similar relative strengths as in W\,33A and NGC\,4945.
The star itself (M star) is most likely not the source of the energetic 
processing. A chance projection with the processed surroundings of 
a young foreground star is possible, but also unlikely. The projected
distance of GC:IRS\,19 to the GC molecular ring is, however, small.
The material in this ring is 
exposed to high UV fluxes from the central cluster, crudely comparable
to the radiation field within the molecular ring in NGC\,4945. It is
hence plausible that energetic processing has created similar ice
properties in both NGC\,4945 and in the GC molecular ring. 
The `pencil beam' line of sight to GC:IRS\,19 may pass through
this processed ring material, explaining the presence of strong
OCN$^-$ absorption in the spectrum of GC:IRS\,19.
The larger beam 14$\arcsec\times$20$\arcsec$ ISO--SWS spectrum of
Sgr\,A$^*$ also shows CO and 
OCN$^-$ ice (Fig.\,\ref{OptDepthFig}; Moneti et al. \cite{Moneti01}). 
The OCN$^-$ column is, however, nearly an order of magnitude smaller 
than in the pencil beam towards GC:IRS\,19. This difference is likely 
due to the large number of background stars within the ISO--SWS beam, 
whose combined line of sight results in the observed low OCN$^-$ optical 
depth. While some of these pencil beams pass through the processed 
GC molecular ring, other pencil beams only sample unprocessed 
foreground material. If the latter line of sight dominates within
the 14$\arcsec\times$20$\arcsec$ ISO--SWS beam, this would be a
natural explanation for the low OCN$^-$ optical depth in the 
Sgr\,A$^*$ spectrum of Moneti et al. (\cite{Moneti01}).

The presence of processed ice in the centers of NGC\,4945 and the
Galaxy leads us to believe that processed ices are one 
characteristic of dense molecular material in star forming nuclear 
regions of galaxies, and can be detected in favorable orientations. 
The amount of processing may then be a measure of the nuclear star 
formation activity and/or geometry.

\section{Conclusions}

Using ISAAC at the VLT to obtain spectra at 2.85--4.10\,$\mu$m
and 4.55--4.80\,$\mu$m, we have detected for the first time `XCN' 
and CO ice in an extragalactic environment: the central region of 
the nearby dusty starburst/AGN galaxy NGC\,4945.

The profile of the solid CO band reveals the importance of thermal
processing of the ice while the prominence of the XCN band attests to
the importance of energetic processing of the ice by FUV radiation,
energetic particles and/or heating. In analogy to the processing of ices
by embedded protostars in our Galaxy, we attribute the processing 
of the ices in the center of NGC\,4945 to ongoing massive star 
formation. 

Our M-band spectrum also shows strong HI\,Pf$\beta$
and H$_2$ 0--0 S(9) line emission and gas phase CO absorption lines.
The HI, H$_2$, gas phase CO and the ices seem to be embedded 
in a rotating molecular disk which is undergoing vigorous star 
formation.

The non-detection of the 3.94\,$\mu$m $[\ion{Si}{ix}]$ coronal line 
in our VLT/ISAAC L-band spectrum is in full agreement with the
very high obscuration towards the AGN derived from mid-infrared
coronal line observations and from the absence of a point-like
source in K-band and N-band images. With the source of the 
obscuration probably close to the AGN for the obscuration to be
effective, the radiation from the AGN cannot be responsible for 
processing the ices.

The obscuration towards the starburst in NGC\,4945 is far higher 
than towards similar starbursts in M\,82 and NGC\,253, also seen 
nearly edge-on. It is unclear whether this is just an on average 
higher absolute absorbing column towards the nuclear region of 
NGC\,4945, or whether there are differences in the properties of 
the absorbing medium. Higher quality M-band observations of 
M\,82 and NGC\,253 are required.

Recently, strong OCN$^-$ absorption has been detected in the spectrum 
of the Galactic center star GC:IRS\,19. The most likely environment 
for the OCN$^-$ absorption is the strongly UV-exposed GC molecular 
ring. The presence of processed ice in the center of NGC\,4945 and 
our Galactic center leads us to believe that processed ice may
be a common characteristic of dense molecular material in star 
forming galactic nuclei.

\begin{acknowledgements} The authors wish to thank Jean Chiar, 
Andrea Moneti and Alessandro Marconi for sharing data with us, 
Fernando Comeron for help in optimising our observing proposal and 
Adwin Boogert, Jean Chiar and Jacqueline Keane for discussions. 
The VLT-ISAAC data were obtained as part of an ESO Service Mode run.
\end{acknowledgements}

\end{document}